\documentstyle[epsfig]{bpl}
\begin{document}
\setcounter{page}{86}

~\\
\bpl  16 April 2009\\
BPL, {\bf 16} (1), pp. 86 - 89, 161013 (2009) \vskip14mm

\title{THE REACH OF THE ATLAS EXPERIMENT IN SUSY PARAMETER SPACE}

\author{J. Dietrich (for the ATLAS collaboration)\\Physikalisches Institut, Universit\"at Freiburg, Germany\\
}
\maketitle

\centerline{(received 7 November 2008; accepted 7 November 2008)}

\abstract{Already with very first data, the ATLAS experiment should be sensitive to a SUSY signal well beyond the regions explored by the Tevatron. We present a detailed study of the ATLAS discovery reach in the parameter space for various SUSY models. The expected uncertainties on the background estimates are taken into account.}\ea

\subsubsection{Introduction}
The search for physics beyond the Standard Model (BSM) is one of the most important goals for the general purpose detector ATLAS at the Large Hadron Collider at CERN. After a short test run in 2008, the LHC, a proton-proton collider with a designed centre-of-mass energy of $\sqrt{s} =14$~TeV, will start in spring 2009 and provide an excellent opportunity to explore such new physics. This paper summarises some strategies of the ATLAS experiment~\cite{ATLAS} to search for direct experimental evidence of Supersymmetry (SUSY) for an integrated luminosity of $1 \rm{fb}^{-1}$. Only a selection of results is shown, a special focus is placed on the discovery reach for inclusive searches for SUSY signatures. 

Searches for SUSY have to deal with models with a relatively large set of free parameters. In this article I will focus on R-parity conserving SUSY particle productions like the mSUGRA, NUHM, GMSB or AMSB models. For R-parity conserving models the strongly interacting SUSY particles (squarks and/or gluinos) are produced in pairs and are unstable. Each will decay via a complicated series of cascade processes into states which include high $p_{T}$ Standard Model particles and the LSPs (lightest SUSY particles). Due to cosmological reasons the LSP is stable, only weakly interacting and will escape the detector unseen.
Therefore SUSY search strategies concentrate on events with large missing transverse energy $E_{T}^{miss}$ and reconstructed particles with large transverse momentum like jets, leptons and photons. The number of jets, leptons or photons strongly depends on the cascade decay of the squarks and/or gluinos and so on the SUSY signal. ATLAS studied various channels with different numbers of jets (1,2,3,4), leptons (0,1,2,3) and  also channels with taus and $b$-jets~\cite{ATLAS}. The goal is to keep the SUSY searches robust and inclusive in order to cover as many signatures and topologies as possible. 
A number of SUSY benchmark points, mostly in the 5-parameter mSUGRA phase space, with a specific choice of model parameters have been selected to study selection cuts. In the following analysis examples the ATLAS point SU3\footnote{$M_0=100$~GeV, $M_{1/2}=300$~GeV, $\tan \beta =6$, $A_0=-300$~GeV, $\mu=+$; $\sigma$ = 28 pb at 14 TeV, ``bulk region''} will be shown. Although mSUGRA imposes particular mass relationships, its phase space is expected to cover a wide variety of signature types expected in the MSSM.
\subsubsection{Inclusive searches for SUSY signals}
In the following I describe a baseline selection as an example for the selection cuts used in the inclusive SUSY searches. Note that for all studies that are presented in this note a data sample of $1\rm{fb}^{-1}$ was assumed. The four jet analysis of ATLAS requires high missing transverse energy $E_{T}^{miss}> 100$~GeV, a transverse sphericity $S_{T}$ cut $>$ 0.2 and at least four jets with $p_T> 50$~GeV and at least one with $p_T>100$~GeV in the event. Accordingly, a $70$~GeV jet trigger combined with an $E_{T}^{miss}> 70$~GeV trigger is required. In the no-lepton search mode events with an isolated high $p_{T}$ ($>20$~GeV) electron or muon are vetoed while for the 1 lepton mode one identified high $p_T$ lepton is required. Moreover for the 1-lepton channel an additional cut on the transverse mass $M_{T}> 100$~GeV, constructed from the identified lepton and the missing transverse energy, is applied.

\begin{figure*}[htb]
\centering
\includegraphics[width=57mm]{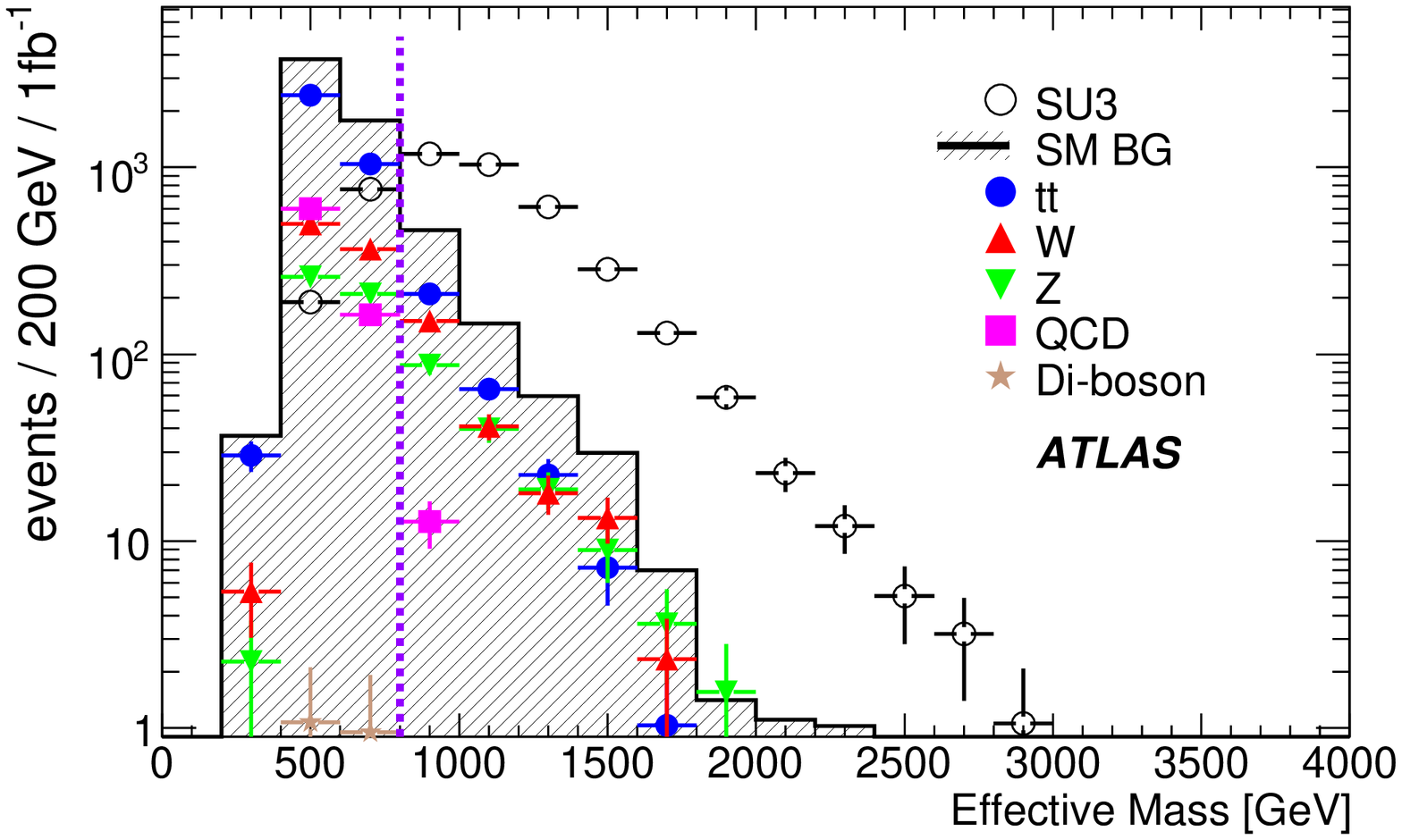}
\includegraphics[width=55mm]{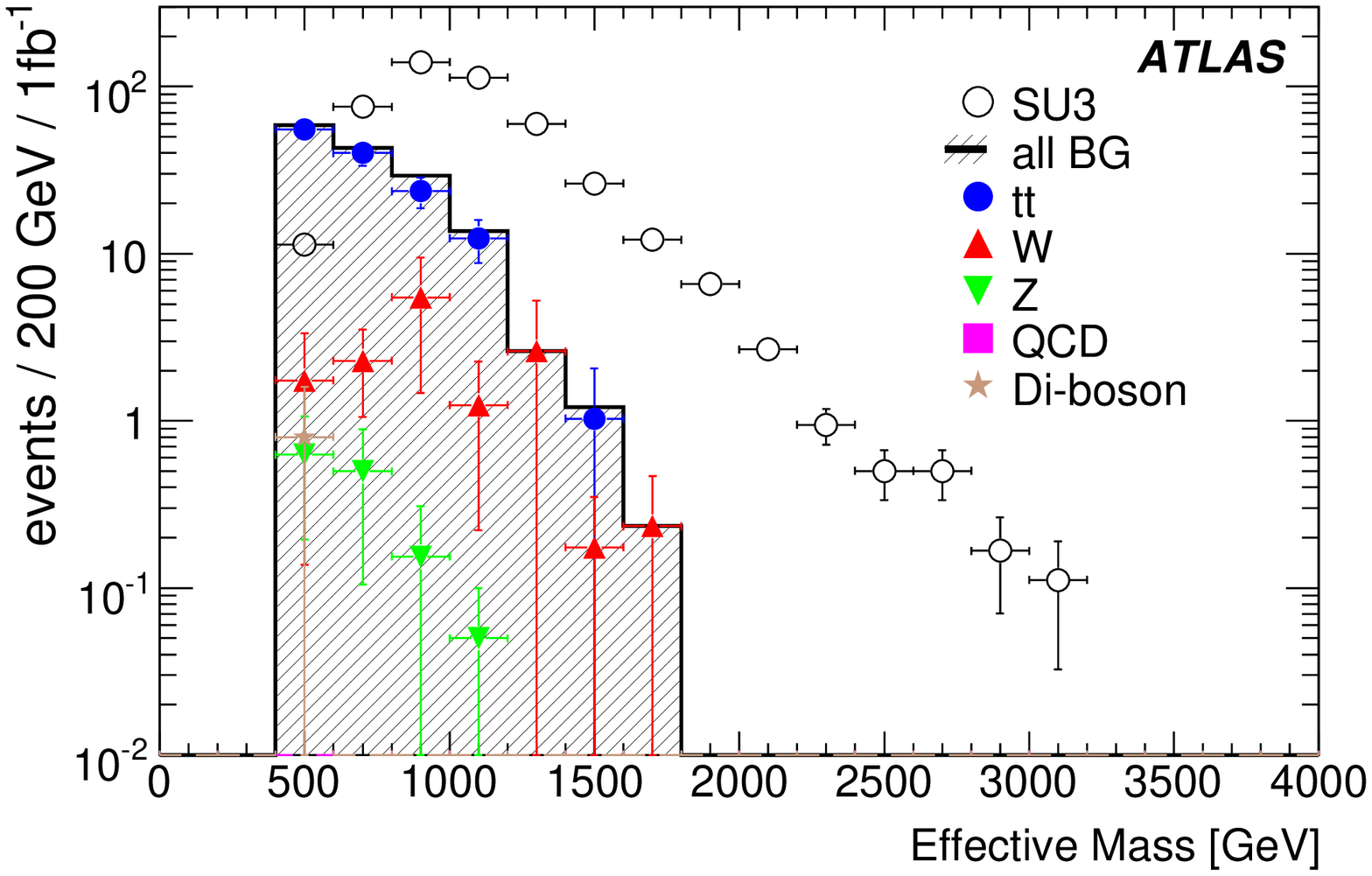}
\put(-98,90){preliminary}
\put(-265,89){preliminary}
\caption{The effective mass $M_{\rm{eff}}$ distribution after final selection for the inclusive jet channel with 0 leptons (left) and 1 lepton (right) at ATLAS.} 
\label{meffplot}
\end{figure*}

To suppress the QCD background the $E_{T}^{miss}$ was required not to point in the direction of jets in $\phi$ and to exceed $0.2\cdot M_{\rm{eff}}$ and $100$~GeV. $M_{\rm{eff}}$ is a discriminating variable between SUSY and background and defined for the 0-lepton mode as the sum of the $p_T$ of the four leading jets and the missing transverse momentum; for the 1-lepton mode the missing transverse momentum of the lepton is also included into $M_{\rm{eff}}$. Figure~\ref{meffplot} shows the effective mass $M_{\rm{eff}}$ distribution of the 4-jet 0-lepton and the 4-jet 1-lepton selection after applying all cuts.
The dominant backgrounds in the 4 jet 0-lepton channel are the $t\bar{t}$, $W$ and $Z\rightarrow \nu \bar{\nu}$ background. In the 1-lepton channel the QCD and $Z$ backgrounds are strongly reduced by the lepton requirement. The main backgrounds in this channel are semileptonic $t\bar{t}$ decays and associated $W$+ jet productions. Further interesting search modes in ATLAS that are not presented in this note include tau leptons or $b$-jets or look for di-leptons and tri-leptons~\cite{ATLAS}. The tri-lepton channel is discussed elsewhere in these proceedings \cite{nonstandard}.

A discovery of new physics can only be claimed when the Standard Model backgrounds are well understood and 
controlled. However for LHC energies Monte Carlo predictions have large uncertainties. 
In order to suppress the dependency on Monte Carlo simulations ATLAS has developed a wide variety of independent strategies to estimate backgrounds from data. 
Large missing transverse energy in QCD events originates mostly from jet mismeasurements. In case of mismeasured jets the 'fake' $E_T^{miss}$ direction will point to one of the jets. A cut in the $\delta\phi = | \phi_{\rm{jet}} - \phi_{\rm{miss}} |$ plane is useful in removing these QCD events by demanding high $\delta\phi$ values. ATLAS has also studied jet smearing techniques to estimate the remaining QCD background as can be found in reference~\cite{ATLAS}.

There is a variety of different methods to estimate the top and $W$ background from data. One is to create a control sample by reversing the 1-lepton mode transverse mass $M_T$ cut. For events where the transverse mass between the lepton and the missing momentum vector is less than $M_W$, the top and $W$ processes are enhanced with respect to SUSY processes~\cite{ATLAS}. 
Assuming the $M_T$ variable uncorrelated with $M_{\rm{eff}}$, the control sample can be used 
to predict the $M_{\rm{eff}}$ distribution in the signal region. 

The $Z \rightarrow \nu \bar{\nu} $+ jets background can be effectively determined from the measurement of $Z \rightarrow e^{+} e^{-} $ and $Z \rightarrow \mu^{+}\mu^{-}$+ jet events. 
Providing sufficient statistics, this method works effectively and provides a nice estimate up to intermediate $M_{\rm{eff}}$.


\subsubsection{Reach in SUSY parameter space}
In order to test a wider range of parameters in the SUSY parameter space several SUSY model parameters were scanned with the goal to develop a search strategy that is covering as wide a subset of the scanned models as possible \cite{ATLAS}.
For each SUSY point the same sets of selection cuts are applied and the significance is calculated. 
The expected 5$\sigma$ discovery reach has been determined for the different SUSY parameter spaces, e.g. the mSUGRA parameter space constrained from measurements including the dark matter density, a model with non-universal higgs masses (NUHM), a mSUGRA model with high $\tan \beta$ and a model with gauge mediated SUSY breaking (GMSB). Note that systematic uncertainties are included to calculate the significance.

\begin{figure*}[htb]
\centering
\includegraphics[width=55mm]{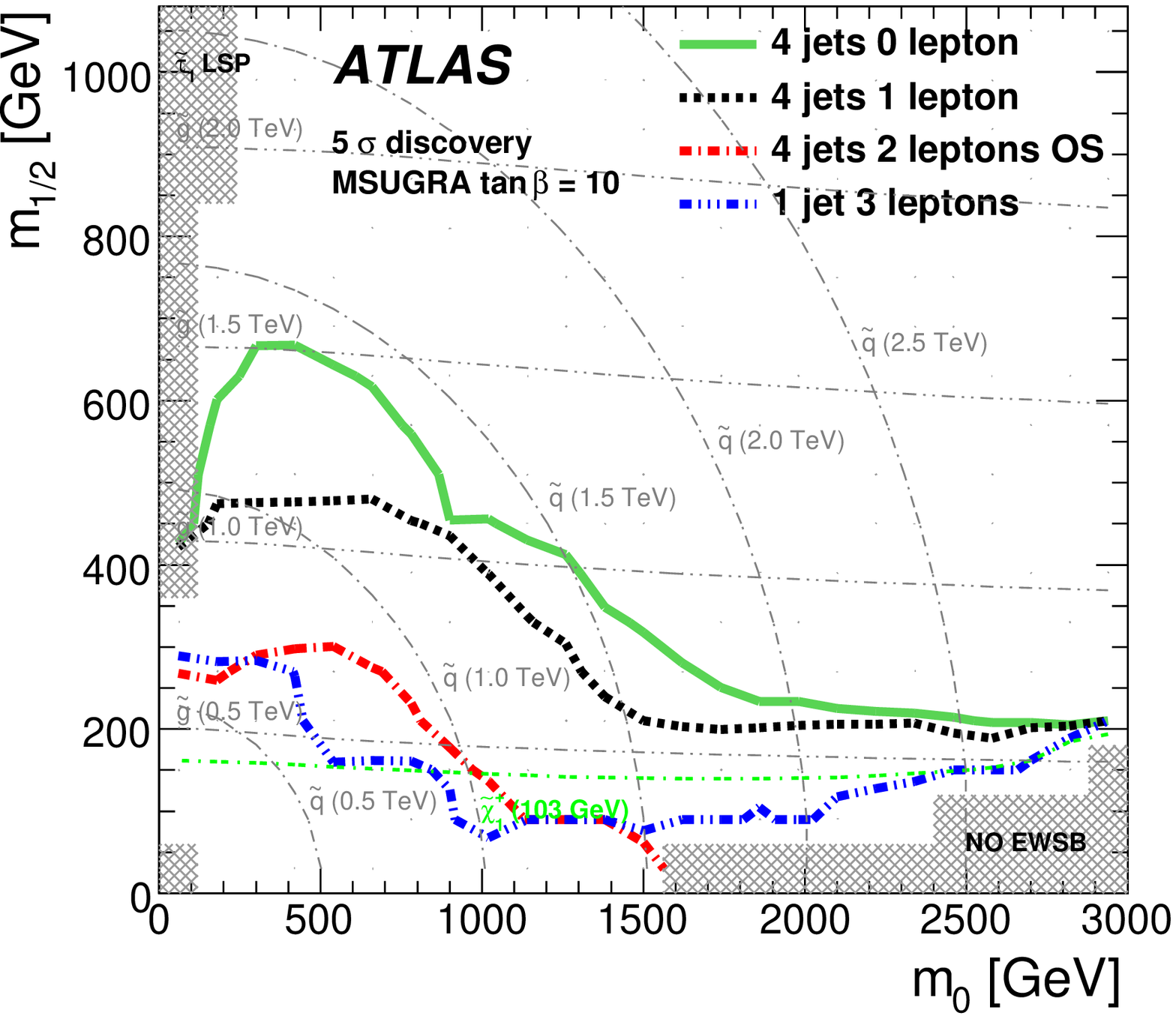}
\includegraphics[width=55mm]{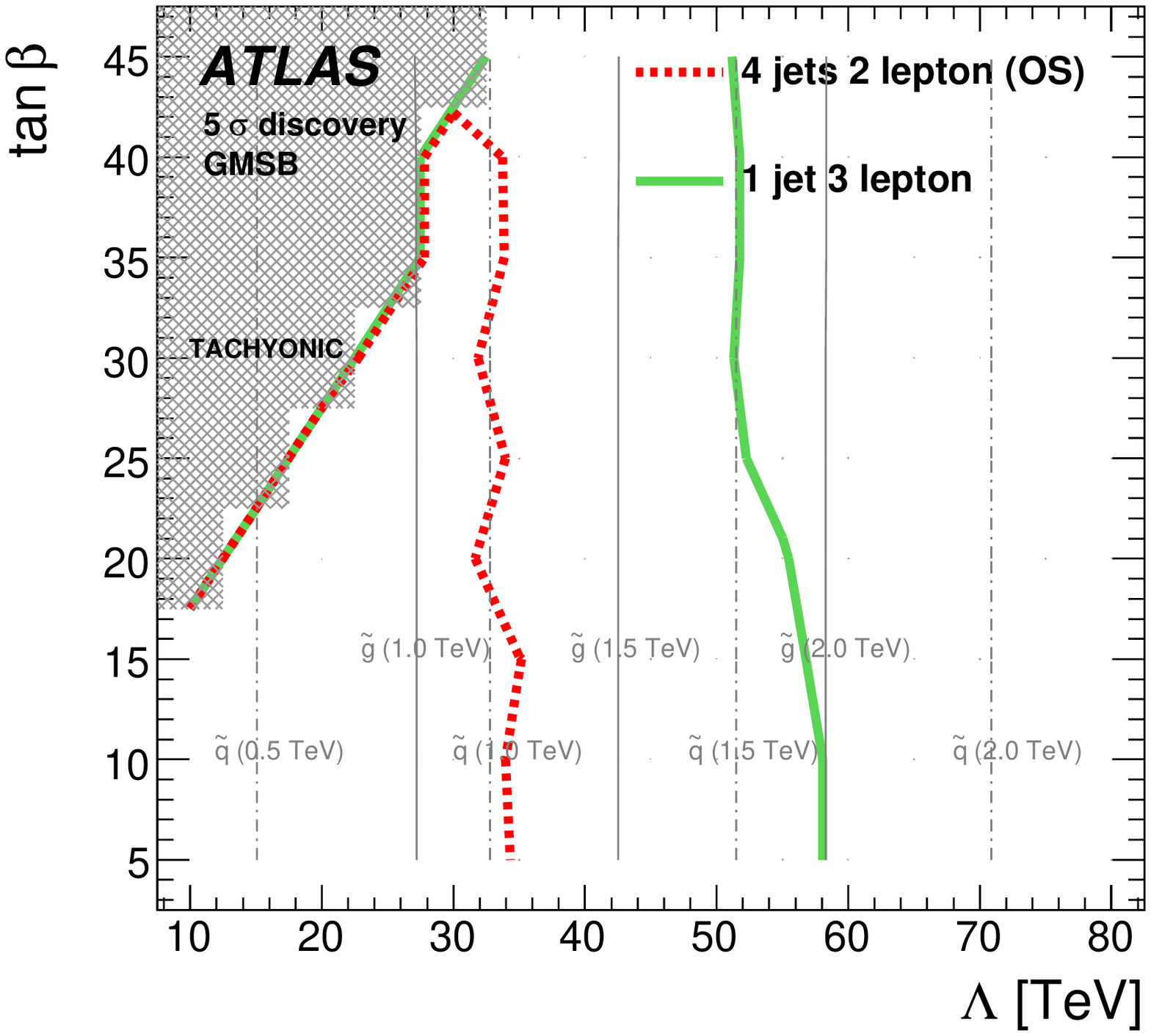}
\put(-240,91){preliminary}
\put(-80,91){preliminary}
\caption{The 1 $\rm{fb}^{-1}$ 5$\sigma$ contours for the ATLAS experiment for the 4 jet analyses and the 1 jet 3 lepton analysis as a function of $M_0$ and $M_{1/2}$ for the mSUGRA model (left) and for the 4 jets 2 leptons and 1 jet 3 leptons channel for the GMSB model (right).}\label{reach}
\end{figure*}

The 5$\sigma$ discovery reach lines in the $M_{0}$-$M_{1/2}$ plane for different channels for the mSUGRA model with $\tan \beta =10$, $A_0=0$~GeV and $\mu=+$ are shown on the left side in figure \ref{reach}. 
The discovery contours for the 4 jets 2 leptons and 1 jet 3 leptons mode for the GMSB model are shown on the right side in figure \ref{reach}. 
\subsubsection{Conclusion}
We are at the threshold of exciting times. 
The results of the scans and the studies with the full simulation show that ATLAS could discover
signals of R-parity conserving SUSY with gluino and squark masses up to 1 TeV after having accumulated and understood the data. 
\subsubsection{Acknowledgments}
I would like to thank the organisers of the ICPP08 for the invitation to present this talk and everyone in the ATLAS Collaboration whose work contributed to this talk.\\
The author acknowledges the support by the Landesstiftung Baden W\"urttemberg and the BMBF.


\begin{thebibliography}{999}   
\bibitem{ATLAS} ATLAS Collaboration, Expected Performance of the ATLAS Experiment, Detector, Trigger and Physics, CERN-OPEN-2008-020, Geneva, 2008, to appear. \\[-7mm]
\bibitem{nonstandard} Searching for new physics in events with three leptons in ATLAS, C. Potter in these proceedings and ATL-PHYS-PROC-2008-049.\\[-7mm]
\end{thebibliography}
\end{document}